\documentclass[sigconf]{acmart}

\AtBeginDocument{%
  }

\setcopyright{acmlicensed}
\copyrightyear{2018}
\acmYear{2018}
\acmDOI{XXXXXXX.XXXXXXX}

\acmConference[Conference 'XX]{The ACM International Conference}{October 21--25, 2018}{California, USA}
\acmISBN{978-1-4503-XXXX-X/18/06}

\usepackage{tabularx}
\usepackage{adjustbox}
\usepackage{multirow}
\usepackage{subfig}
\usepackage{fontawesome}
\usepackage{xspace}
\usepackage{graphicx}

\usepackage{tcolorbox}

\newcommand{\dataset}{\texttt{SynDL}\xspace}
\newcommand{\TitleFigure}{\includegraphics[scale=0.12]{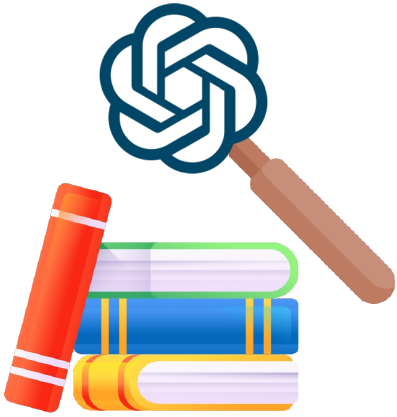}\xspace}

\begin{document}

% \title[\dataset: A Large-Scale Synthetic Test Collection]{\includegraphics[scale=0.12]{figs/DeepJudge-logo.pdf} \dataset: A Large-Scale Synthetic Test Collection}
% \posttitle{
%   \\[\bigskipamount]\includegraphics[scale=0.12]{figs/DeepJudge-logo.pdf}
% }
\title[\dataset: A Large-Scale Synthetic Test Collection]{\TitleFigure \dataset: A Large-Scale Synthetic Test Collection \\ for Passage Retrieval}

% \title[\dataset: A Large-Scale Synthetic Test Collection]{\includegraphics[scale=0.12]{figs/DeepJudge-logo.pdf} \dataset \\ A Large-Scale Synthetic Test Collection for Passage Retrieval}

 % for Passage Retrieval
\author{
Hossein A.~Rahmani
}
\orcid{0000-0002-2779-4942} 
\affiliation{%
        \institution{University College London}
        \city{London}
        \country{UK}
}
\email{hossein.rahmani.22@ucl.ac.uk}

\author{Xi Wang}
\orcid{0000-0001-5936-9919} 
\affiliation{%
        \institution{University of Sheffield}
        \city{Sheffield}
        \country{UK}
}
\email{xi.wang@sheffield.ac.uk}

\author{Emine Yilmaz}
\orcid{0000-0003-4734-4532} 
\affiliation{%
        \institution{University College London \\ Alan Turing Institute \& Amazon}
        \city{London}
        \country{UK}
}
\email{emine.yilmaz@ucl.ac.uk}

\author{Nick Craswell}
\orcid{0000-0002-9351-8137} 
\affiliation{%
        \institution{Microsoft}
        \city{Seattle}
        \country{US}
}
\email{nickcr@microsoft.com}

\author{Bhaskar Mitra}
\orcid{0000-0002-5270-5550} 
\affiliation{%
        \institution{Microsoft}
        \city{Montréal}
        \country{Canada}
}
\email{bmitra@microsoft.com}

\author{Paul Thomas}
\orcid{0000-0003-2425-3136} 
\affiliation{%
        \institution{Microsoft}
        \city{Adelaide}
        \country{Australia}
}
\email{pathom@microsoft.com}

\renewcommand{\shortauthors}{H.~A.~Rahmani et al.}
\newcommand{\xw}[1]{\textcolor{blue}{#1}}
\newcommand{\saeed}[1]{\textcolor{red}{#1}}
\begin{abstract}
  Large-scale test collections play a crucial role in Information Retrieval (IR) research. However, according to the Cranfield paradigm and the research into publicly available datasets, the existing information retrieval research studies are commonly developed on small-scale datasets that rely on human assessors for relevance judgments — a time-intensive and expensive process. Recent studies have shown the strong capability of Large Language Models (LLMs) in producing reliable relevance judgments with human accuracy but at a greatly reduced cost. In this paper, to address the missing large-scale ad-hoc document retrieval dataset, we extend the TREC Deep Learning Track (DL) test collection via additional language model synthetic labels to enable researchers to test and evaluate their search systems at a large scale. Specifically, such a test collection includes more than 1,900 test queries from the previous years of tracks. We compare system evaluation with past human labels from past years and find that our synthetically created large-scale test collection can lead to highly correlated system rankings.
\end{abstract}







\maketitle

\section{Introduction}
\label{sec:intro}
Text retrieval approaches identify complete documents relevant to a query. Relevance can be computed based on the similarity of the query and document as determined by comparing the query to words and passages in documents \cite{marcin1997passage}. Alternatively, due to the likelihood of including non-relevant or redundant information in a document and the efficiency of locating relevant information in documents, passage retrieval has become a common research task in information retrieval with the development of many passage retrieval models \cite{wade2005passage}. Meanwhile, the introduction of effective passage retrieval solutions strongly ties to a well-rounded evaluation. Many existing evaluation operations are commonly instructed by the Cranfield paradigm \cite{cleverdon1960aslib} using a test collection to determine the performance of an information retrieval system. A basic test collection needs to comprise a large set of documents or passages, a set of information needs in plain text, and the corresponding relevance judgments for every document or passage when referring to each information need. Many known and commonly investigated test collections include MS MARCO \cite{tri2016msmarco}, the collections released by years of organisation of evaluation campaigns like TREC (e.g., TREC Deep Learning Tracks \cite{craswell2020overview,craswell2024overview}), and workshops or conferences like CLEF and NTCIR.     

However, even though there are many available test collections \cite{macavaney2021simplified}, it has been a common concern in the information retrieval community about the shortage of a large-scale test collection for modelling the complex relationships between queries and documents and developing advanced passage and document ranking approaches \cite{faggioli2023perspectives}. Indeed, using MS MARCO as a typical test collection example, it has over 1M of questions that can act as queries. However, for each query, only an average of 10 passages may contain the answer to the query, leaving about 8.8M passages as non-relevant \cite{tri2016msmarco}. Similarly, for the test collection of TREC Deep Learning (2023) \cite{craswell2024overview}, even though it has richer labels about the query to passages in different relevance levels (related, highly relevant, perfectly relevant), we still observe a low number of queries (i.e, 82) for evaluation, which has been the highest since the tracks from 2019. Hence, it is difficult for a model to capture the complex relationships when modelling the relevance of a query to a large passage corpus, especially in the initial ranking stage. 

On the other hand, over recent years, the rapid advancement of machine learning techniques, especially with the introduction of large language models capable of promising natural language comprehension and generation ability, has greatly updated the research development strategies in the information retrieval community. Some examples are the growing research outputs on dense passage retrieval techniques \cite{zhao2024dense}, instructed language model for user-intent aware retrieval \cite{asai2023task}, and Query2Doc \cite{wang2023query2doc} that generates pseudo documents based on a query by prompting language models. In particular, following the effectiveness of natural language generation via large language models, we see the potential of language models in making judgments about the relevance between queries and documents \cite{rahmani2024llm4eval,rahmani2024report,rahmani2024llmjudge}. Indeed, as discussed in \cite{rahmani2024synthetic}, a high correlation between using human and LLM judgments when assessing system rankings has been observed, which encourages the introduction of a large-scale test collection to domains in need. Hence, in this paper, we aim to contribute to the development of a high-quality large-scale passage retrieval test collection with the use of large language models. It is worth noting that large language models could act as a ranking model directly. However, it is essential to adopt large-scale language models to ensure satisfactory performance \cite{faggioli2023perspectives} and its use can be time and economically costly to the retrieval process. In this study, we ground our development by extending the test collections from the five years of the TREC Deep Learning (DL) tracks using large language models to distil the relevance assessment. We refer to the developed passage retrieval test collection with the name of \dataset, its release and many associated baseline approaches from years of TREC DL submissions aim to support in addressing the following research challenges in the community: 

 \begin{itemize}
     \item \textbf{Deep Relevance Assessment}: The existing passage retrieval test collection often provides few relevance labels on documents for each query, which results in shallow evaluation.   
     \item \textbf{Diverse Evaluation}: Many passage test collections, like the ones used in TREC DL tracks, use a small number of queries and limit the evaluation to a small set of test query samples.
     \item \textbf{Rich Baselines}: The inclusion of a small list of baselines often ignores the rich insightful comparisons while introducing novel techniques.
     \item \textbf{Synthetic Query Analysis}: Existing test collections do not enable extensive analysis into the case of comparing synthetic queries and human-provided queries with deep query relevance labels.
 \end{itemize}
 
In this study, with the notice of the above challenges, we first provide a comprehensive and detailed description of the \dataset test collection. In addition, we augment the introduction of this test collection with extensive evaluations on the alignment of using LLM judgments to human judgments, the comparison of the difference of the resulting system ranking orders and the potential bias effect that might introduced by the use of LLM judgments. With the in-depth evaluation and analysis, we show the high quality of our test collection in providing aligned passage retrieval system rankings to human assessors with ``deep and wide'' relevance labels.
\section{Related Work}
\label{sec:relatedwork}
By following the Cranfield paradigm \cite{cleverdon1960aslib}, it is a common practice in the information retrieval community to
evaluate the performance of a passage ranking model on benchmark test collections. The development of passage ranking test collections regularly happens with the organisation of many evaluation campaigns, such as TREC and CLEF. For example, the MS MARCO dataset \cite{tri2016msmarco} was introduced with the inclusion of various components, such as queries, passages, answers and documents. In particular, to be used for the passage ranking task, each query is associated with an average of 10 passages, which might be relevant to provide the answer. Later, with the annual organisation of the TREC DL tracks starting from 2019 \cite{craswell2020overview}, extensive additional human labelling was conducted and released each year to often the relevance of between a set of queries to passages to a different extent. However, these test collections are all limited to having a small number of queries and no more than 100 (82 on DL-23 at maximum). Similarly, the ClueWeb series corpus (ClueWeb09, 12, 22) was introduced with from roughly 1 billion web pages of ClueWeb09 to 10 billion web pages of ClueWeb22. Then, the TREC Web Tracks \cite{clarke2009overview} were organised and relevance assessors were recruited to judge if a web document could satisfy the information needs expressed by queries from logs of commercial search engines. It's noteworthy that for the diversity search task, the test collections of the TREC Web Tracks rely on a rich set of relevant documents for diversity evaluation. However, due to the high cost of human assessment, only 50 queries are included in the test collections for evaluation. There are many other document or passage retrieval test collections that are publicly available, such as BEIR \cite{thakur2021beir}, and TREC Disks 4 and 5 that were used in the TREC robust 2004 track \cite{voorhees2005trec}. However, either the shortage of rich queries or the shallow set of relevant documents to queries is a common observation among the existing test collections, especially for the passage retrieval task. Meanwhile, an agreement on the high correlation of system performance evaluation using human assessments and LLM judgments has also been extensively discussed in the literature \cite{faggioli2023perspectives,thomas2023large}. Specifically, the use of LLM judgments could vary from an assistant role for human annotation, like in \cite{jayasinghe2014improving} to an automatic annotator. The effectiveness of LLM judgments would require in-depth analysis and investigations. However, thus far, it is still missing a large-scale passage ranking test collection that harvests the contributions of both human and LLM annotators, which leads to the development and release of our \dataset test collection in this paper.
\section{\dataset Test Collection Development}
\label{sec:collection}
With a focus on the task of passage ranking, we aim to extend the popular test collections of TREC Deep Learning tracks and develop a large-scale test collection, named \dataset, by leveraging LLM judgments to mitigate the discussed research challenges caused by the shortage of a test collection with diversified queries and deep document relevance labels. To illustrate the test collection development process, we first describe the base test collections sourced from the TREC Deep Learning tracks, then followed by the test collection extension strategy with the use of LLM judgments. 

The TREC Deep Learning (DL) Track is an initiative organized by the National Institute of Standards and Technology (NIST) to advance the state-of-the-art in information retrieval (IR) and related tasks using deep learning techniques. This track focuses on evaluating the performance of deep learning methods on large-scale datasets and encourages the development of new models and techniques in this domain. The organisation of TREC DL Track was initiated in 2019 \cite{craswell2020overview} and has its final edition in 2023. It has a main focus on two information retrieval tasks, document retrieval and passage retrieval. In particular, each task uses labels provided by human assessors that justify if a passage can answer a given query from the MS MARCO dataset. Note that, in the last run of DL-23, ``synthetic queries'' are also included in the test collection for non-official evaluation to gain additional insights when compared to the official human evaluations. In Table~\ref{tbl:dl-statistics}, we present a statistical summary of the TREC Deep Learning test collections over the five years of runs. It is noticeable that, on average, all the test collections rely on a small set of test queries but a reasonable size of relevant documents for performance assessment. However, it is known that the use of small-size test samples could result in inconsistent observation when compared to the use of complex and diversified test samples \cite{sanderson2010test}, especially when ``wide and shallow'' can outweigh ``deep and narrow'' test collections empirically \cite{carterette2008evaluation}. 

In this paper, to improve the resources from existing TREC Deep learning track runs, we propose to leverage the advanced ability of language models to comprehend natural language and assess the relevance between queries and documents \cite{faggioli2023perspectives}. Specifically, the development of the extended \dataset test collection is organised in three stages: (1) Initial Query Assemble, (2) Assessment Pool Generation and (3) Automatic judgment with LLM. We provide the corresponding descriptions as follows:

\begin{table}
    \centering
    \caption{\dataset dataset statistics}
    \label{tbl:dl-statistics}
    \begin{adjustbox}{max width=\columnwidth}
        \begin{tabular}{lccccc|c}
            \toprule
            \textbf{Data} & \textbf{DL-19} & \textbf{DL-20} & \textbf{DL-21} & \textbf{DL-22} & \textbf{DL-23} & \textbf{\dataset}\\
            \midrule
             TREC (Judged) Queries  & 43    & 54     & 53     & 76       & 82     & \textbf{1,988} \\
             TREC (Initial) Queries & 200   & 200    & 477    & 500      & 700    & \textbf{1,988} \\
             TREC Qrels             & 9,260 & 11,386 & 10,828 & 386,416  & 22,327 & \textbf{637,063} \\
             TREC Qrels/Query       & 215.3 & 210.9  & 204.3  & 5,084.4  & 272.2  & \textbf{320.45} \\
             TREC Docs              & 8.8M  & 8.8M   & 138M   & 138M     & 138M   & \textbf{146.8M} \\
             \midrule
             Irrelevant (0)         & 5,158 & 7,780 & 4,338 & 286,459 & 13,866 & 369,567 \\
             Related (1)            & 1,601 & 1,940 & 3,063 & 52,218  & 4,372  & 126,406 \\
             Highly relevant (2)    & 1,804 & 1,020 & 2,341 & 46,080  & 2,259  & 86,162 \\
             Perfectly relevant (3) & 697   & 646   & 1,086 & 1,659   & 1,830  & 54,928 \\
            \bottomrule
        \end{tabular}
    \end{adjustbox}
\end{table}

\begin{enumerate}
    \item \textbf{Initial Query Assemble:} For each test collection resource of Deep Learning tracks, it is associated with a set of initial queries, which was meant to be used for human annotators to assess the document relevance for a full set of initial queries. However, only a portion of queries were selected by the human assessor to provide relevant judgments. For example, regarding the test collection sourced DL-19, 200 initial queries were provided but the human assessors left 157 queries unlabelled. Although the following years of runs increased initial queries up to 700, the selected queries for assessment remained on a small scale. To increase the diversity of the queries included in the test collection, we aggregate all initial queries, with a size of 1,988, sourced from the five runs of Deep learning tracks as the initial query inputs for the use of later LLM judgment. It is worth noting that the initial queries of DL-23 also include 500 synthetic queries generated by GPT-4 and T5 models with 250 synthetic queries each \cite{rahmani2024synthetic}. The inclusion of synthetic queries can allow additional bias investigation study of LLM judgments on synthetic queries.   
    \item \textbf{Assessment Pool Generation:} With the collected initial queries, we follow the common practice in TREC passage retrieval system assessment to prepare the passage pool. In a TREC passage retrieval evaluation, each submission is required to submit a ranked list of passages for each query. Then, the evaluation will be made by selectively considering varied depths of ranked passages with a set of evaluation metrics, such as NDCG@5 and NDCG@10. For the development of diversified relevant passages in our test collection, we embrace the use of the rich submissions among the five runs to collect a depth-10 pool with good coverage of passages in high relevant probabilities. Overall, we use 37, 59, 63, 100 and 35 submissions corresponding to the runs from DL-19 to DL-23. Note that, we will also include these submissions as baselines with their full descriptions in our GitHub repository. After removing the overlapped passages, we obtain a full set of $637,063$ query-passage pairs for relevance assessment. On average, each query is associated with $320.45$ passages for relevance annotation.     
    \item \textbf{Automatic Judgment with LLM:} After having the query-passage pairs ready, we start the annotation of these inputs via large language models. Specifically, with the recently verified advance of GPT-4 \cite{achiam2023gpt} in many natural language tasks, we use GPT-4 for this annotation task with a devised prompt and ask the model to provide annotations in a high granularity (i.e., related, highly relevant and perfectly relevant). It is interesting to observe that human annotators are more sensitive to giving perfectly relevant judgments, while GPT-4 give about equal labels of highly relevant and perfectly relevant. Due to the space limit, we include the used prompt in our \href{https://rahmanidashti.github.io/SynDL/}{GitHub}\footnote{\url{https://rahmanidashti.github.io/SynDL/}} repository for reproducible generation. 
    % Regarding the annotation cost, ... 
\end{enumerate}

After the annotation with LLM judgments, we receive a large-scale test collection with 637,063 query-passage relevance labels and a rich set of queries (1,988). However, as a test collection, it is essential to evaluate the quality of the generated LLM judgments. Hence, we also conduct extensive analysis on the generated test collection for quality evaluation.

\begin{figure}
    \centering
    \subfloat[NDCG@10\label{fig:dl-19-ndcg10}]
    {
        {
            \includegraphics[scale=0.38]{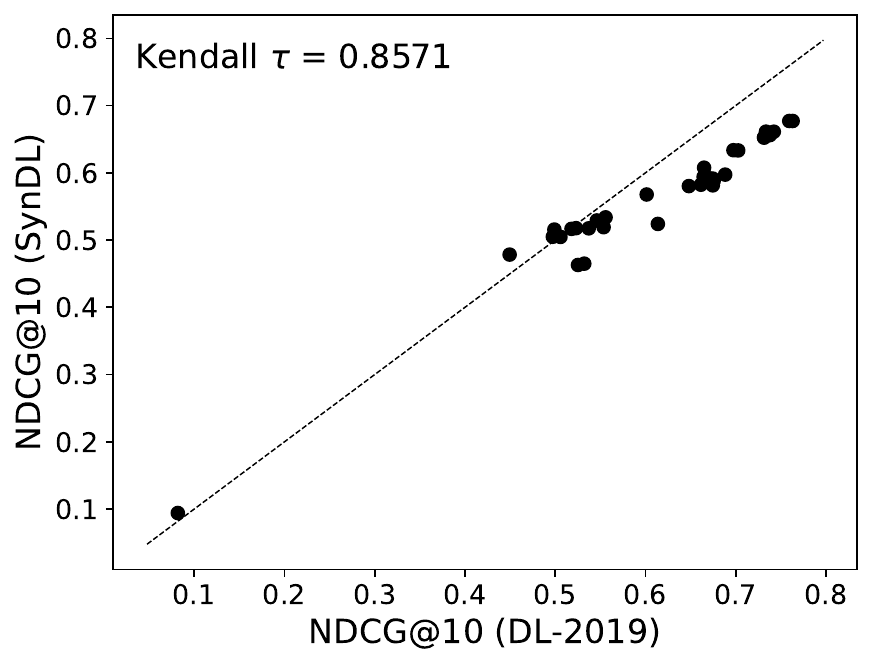}
        }
    }%
    \quad
    \subfloat[NDCG@100\label{fig:dl-19-ndcg100}]
    {
        {
            \includegraphics[scale=0.38]{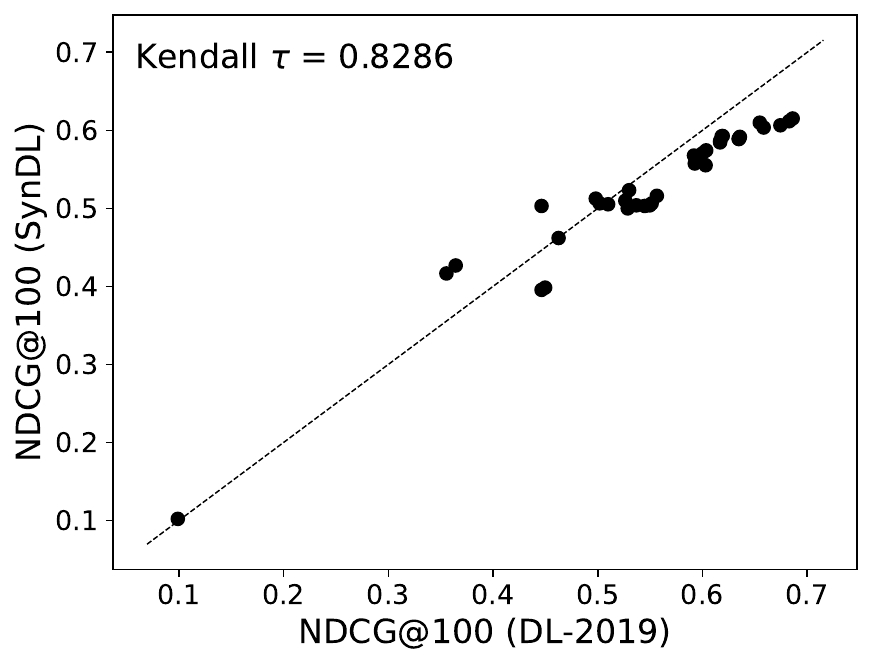}
        }
    }%
    \caption{System Ranking correlation test between two test collections (DL-19 and \dataset).}%
    \label{fig:dl-19}%
\end{figure}

\section{Resource Evaluation}
\begin{table}
    \centering
    \caption{Five top-performing submission runs on \dataset}
    \label{tbl:numrels_perquery}
    \begin{adjustbox}{max width=\columnwidth}
        \begin{tabular}{lccccc}
            \toprule
            \textbf{Run} &  Rank (DL-23) & \textbf{Run type} & \textbf{NDCG@10} & \textbf{NDCG@100} & \textbf{AP} \\
            \midrule
             naverloo-rgpt4    &  1  & prompt & 0.9060 & 0.7841 & 0.5628 \\
             naverloo-frgpt4   &  2  & prompt & 0.9007 & 0.7841 & 0.5651 \\
             naverloo\_fs\_RR\_duo & 3 & prompt & 0.8849 & 0.7782 & 0.5590 \\
            cip\_run\_2       &  4  & prompt & 0.8671 & 0.7101 & 0.4866 \\
             cip\_run\_1       &  5  & prompt & 0.8671 & 0.7101 & 0.4867 \\
             % \midrule
            \bottomrule
        \end{tabular}
    \end{adjustbox}
\end{table}

\begin{figure}
    \centering
    \includegraphics[scale=0.38]{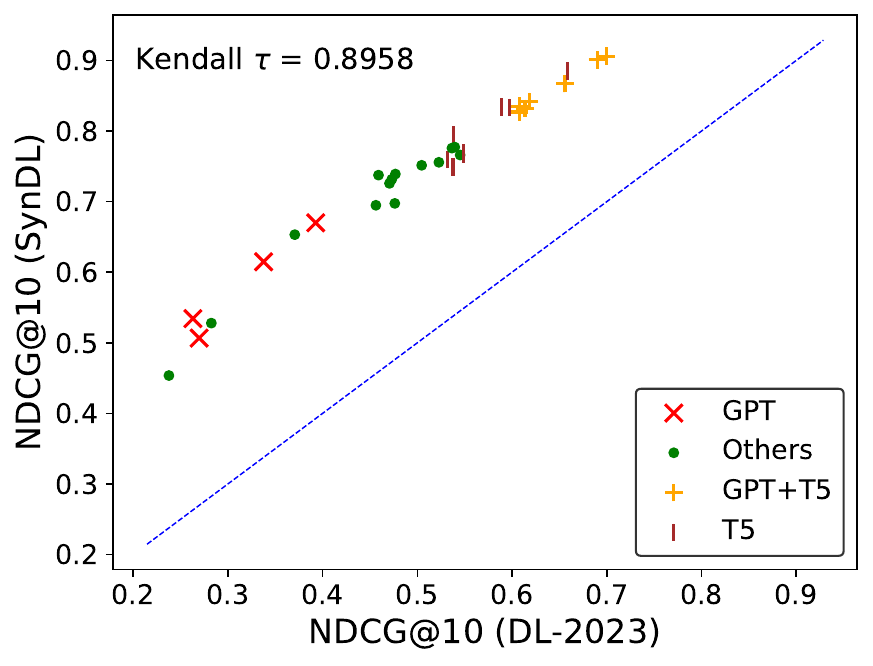}
    \caption{Scatter plots of the effectiveness of DL-23 runs based on \dataset synthetic queries vs.~DL-23 test collection to analyse the bias towards systems using the same language model as the one used in synthetic query construction.}
    \label{fig:enter-label}
\end{figure}

To effectively evaluate our \dataset test collection, we follow the evaluation setups in \cite{faggioli2023perspectives,rahmani2024synthetic}, which use the correlation test on the system ranking when evaluated using human judgments and LLM judgments. In particular, we compare the performance of systems that were submitted to the five runs of deep learning tracks. Note that, due to the space limit, we only present the correlation test results on DL-19 in this paper and we observe similar results across the comparison on all TREC DL test collections. The rest of the test results will be made available in our \href{https://github.com/rahmanidashti/}{GitHub} repository for a complete comparison. With the evaluation of the performance of 37 systems that were submitted to DL-19 for official judgment, we compare the ranking difference of these submissions between the use of human assessments and synthetic relevance judgments in our \dataset test collection. Figure~\ref{fig:dl-19} shows the evaluated correlation via Kendall rank correlation coefficients when evaluated with NDCG in two depths (@10 and @100). The line of $y = x$ is also included for comparison. According to the value of the correlation test results, we observe a high system ordering agreement using the two test collections with Kendall's $\tau$ = 0.8571 and 0.8286 for NDCG@10 and @100, respectively. 

In addition to the system ordering agreement evaluation, we also include another comparative study about the agreement on the top-ranked passage ranking systems. The purpose of this evaluation is to observe if the top-ranked systems can also be observed when evaluated on our \dataset test collection, which can further justify the correlation of the made LLM judgments with the human assessors. In Table~\ref{tbl:numrels_perquery}, we present the evaluation results of top-ranked submissions to the DL-23 on our \dataset test collection. First, regarding the rank difference, we can see that the top 5 systems remain the best-performing passage rankers among the rest of the submissions to DL-23. In particular, the two ranking orders are identical if we consider the NDCG@10 and NDCG@100 measures.

Moreover, with the inclusion of synthetic queries in our \dataset test collection, we further conduct a bias analysis to examine if our test collection would favour systems also using the same or similar language models. As discussed in \cite{liu2023g}, there is a potential bias towards LLM-generated text when using LLM for evaluation. Hence, to explore this bias effect, we first categorise the submissions to the DL-23 into four categories according to whether they are based on GPT (\textbf{$\times$}), T5(\textbf{|}), GPT + T5 (\textbf{+}) or others ($\boldsymbol{\cdot}$). Figure \ref{fig:enter-label} shows the system order agreement between the use of our \dataset and DL-23 with highlighted different types of systems. We observe that a high agreement can still be observed between human assessment and language model judgments in this case. GPT-based systems do not get higher ranks when evaluated with GPT-4 generated relevance judgments in our \dataset test collection. 

Overall, we experimentally verify that our \dataset test collection is of a high quality, which not only exhibits a high agreement with the human assessors across the comparison to multiple sets of test collections but also shows a robust evaluation outcome when evaluated on the potential bias about using the identical language models.
\section{Discussions and Conclusions}
In this paper, we summarise the construction of a large-scale test collection (\dataset) with LLM-based relevance judgment for passage retrieval, which is developed based on the test collections from the five runs of TREC Deep Learning tracks. The resulting test collection, \dataset, covers a rich set of queries with deep relevance labels on passages. After conducting a thorough quality evaluation of \dataset, we observe a high agreement between \dataset and every TREC DL test collection on system ordering. In addition, we also highlighted that \dataset with language model relevance judgment does not favour language model approaches according to our conducted experiments. 

Recall the observed research challenges for the passage retrieval task. We conclude that our \dataset test collection is promising in providing deep relevance assessment with rich relevance labels and a diverse set of queries. In addition, with the inclusion of passage retrieval systems that were submitted to the TREC DL tracks, we enable the comparison over a rich set of baseline approaches. Moreover, by comparing the research findings on the real and synthetic queries, \dataset also allows extensive research studies to evaluate passage retrieval systems on different types of queries. While our preliminary analysis showed the promising value of our \dataset test collection, we also see the many possibilities of research contributions with its release, such as the transfer learning of using models pre-trained on our test collection, the re-visit of many existing passage retrieval approaches and the development of generalisable passage retrieval techniques on diverse queries.

\begin{acks}
This work is supported by the Engineering and Physical Sciences Research Council [EP/S021566/1], the EPSRC Fellowship titled ``Task Based Information Retrieval'' [EP/P024289/1], the Turing Fellowship scheme.
\end{acks}

\bibliographystyle{ACM-Reference-Format}
\bibliography{bib}


\begin{thebibliography}{24}

%%% the Web address.

\ifx \showCODEN    \undefined \def \showCODEN     #1{\unskip}     \fi
\ifx \showDOI      \undefined \def \showDOI       #1{#1}\fi
\ifx \showISBNx    \undefined \def \showISBNx     #1{\unskip}     \fi
\ifx \showISBNxiii \undefined \def \showISBNxiii  #1{\unskip}     \fi
\ifx \showISSN     \undefined \def \showISSN      #1{\unskip}     \fi
\ifx \showLCCN     \undefined \def \showLCCN      #1{\unskip}     \fi
\ifx \shownote     \undefined \def \shownote      #1{#1}          \fi
\ifx \showarticletitle \undefined \def \showarticletitle #1{#1}   \fi
\ifx \showURL      \undefined \def \showURL       {\relax}        \fi
% invisible to TeX
\providecommand\bibfield[2]{#2}
\providecommand\bibinfo[2]{#2}
\providecommand\natexlab[1]{#1}
\providecommand\showeprint[2][]{arXiv:#2}

\bibitem[Achiam et~al\mbox{.}(2023)]%
        {achiam2023gpt}
\bibfield{author}{\bibinfo{person}{Josh Achiam}, \bibinfo{person}{Steven Adler}, \bibinfo{person}{Sandhini Agarwal}, \bibinfo{person}{Lama Ahmad}, \bibinfo{person}{Ilge Akkaya}, \bibinfo{person}{Florencia~Leoni Aleman}, \bibinfo{person}{Diogo Almeida}, \bibinfo{person}{Janko Altenschmidt}, \bibinfo{person}{Sam Altman}, \bibinfo{person}{Shyamal Anadkat}, {et~al\mbox{.}}} \bibinfo{year}{2023}\natexlab{}.
\newblock \showarticletitle{Gpt-4 technical report}.
\newblock \bibinfo{journal}{\emph{arXiv preprint arXiv:2303.08774}} (\bibinfo{year}{2023}).
\newblock


\bibitem[Asai et~al\mbox{.}(2023)]%
        {asai2023task}
\bibfield{author}{\bibinfo{person}{Akari Asai}, \bibinfo{person}{Timo Schick}, \bibinfo{person}{Patrick Lewis}, \bibinfo{person}{Xilun Chen}, \bibinfo{person}{Gautier Izacard}, \bibinfo{person}{Sebastian Riedel}, \bibinfo{person}{Hannaneh Hajishirzi}, {and} \bibinfo{person}{Wen-tau Yih}.} \bibinfo{year}{2023}\natexlab{}.
\newblock \showarticletitle{Task-aware Retrieval with Instructions}. In \bibinfo{booktitle}{\emph{Findings of the Association for Computational Linguistics: ACL 2023}}. \bibinfo{pages}{3650--3675}.
\newblock


\bibitem[Carterette et~al\mbox{.}(2008)]%
        {carterette2008evaluation}
\bibfield{author}{\bibinfo{person}{Ben Carterette}, \bibinfo{person}{Virgil Pavlu}, \bibinfo{person}{Evangelos Kanoulas}, \bibinfo{person}{Javed~A Aslam}, {and} \bibinfo{person}{James Allan}.} \bibinfo{year}{2008}\natexlab{}.
\newblock \showarticletitle{Evaluation over thousands of queries}. In \bibinfo{booktitle}{\emph{Proceedings of the 31st annual international ACM SIGIR conference on Research and development in information retrieval}}. \bibinfo{pages}{651--658}.
\newblock


\bibitem[Clarke et~al\mbox{.}(2009)]%
        {clarke2009overview}
\bibfield{author}{\bibinfo{person}{Charles~LA Clarke}, \bibinfo{person}{Nick Craswell}, {and} \bibinfo{person}{Ian Soboroff}.} \bibinfo{year}{2009}\natexlab{}.
\newblock \showarticletitle{Overview of the TREC 2009 Web Track.}. In \bibinfo{booktitle}{\emph{Trec}}, Vol.~\bibinfo{volume}{9}. \bibinfo{pages}{20--29}.
\newblock


\bibitem[Cleverdon(1960)]%
        {cleverdon1960aslib}
\bibfield{author}{\bibinfo{person}{Cyril~W Cleverdon}.} \bibinfo{year}{1960}\natexlab{}.
\newblock \showarticletitle{The aslib cranfield research project on the comparative efficiency of indexing systems}. In \bibinfo{booktitle}{\emph{Aslib Proceedings}}, Vol.~\bibinfo{volume}{12}. \bibinfo{pages}{421--431}.
\newblock


\bibitem[Craswell et~al\mbox{.}(2020)]%
        {craswell2020overview}
\bibfield{author}{\bibinfo{person}{Nick Craswell}, \bibinfo{person}{Bhaskar Mitra}, \bibinfo{person}{Emine Yilmaz}, \bibinfo{person}{Daniel Campos}, {and} \bibinfo{person}{Ellen~M Voorhees}.} \bibinfo{year}{2020}\natexlab{}.
\newblock \showarticletitle{Overview of the TREC 2019 deep learning track}. In \bibinfo{booktitle}{\emph{Trec}}.
\newblock


\bibitem[Craswell et~al\mbox{.}(2024)]%
        {craswell2024overview}
\bibfield{author}{\bibinfo{person}{Nick Craswell}, \bibinfo{person}{Bhaskar Mitra}, \bibinfo{person}{Emine Yilmaz}, \bibinfo{person}{Hossein~A. Rahmani}, \bibinfo{person}{Daniel Campos}, \bibinfo{person}{Jimmy Lin}, \bibinfo{person}{Ellen~M. Voorhees}, {and} \bibinfo{person}{Ian Soboroff}.} \bibinfo{year}{2024}\natexlab{}.
\newblock \showarticletitle{Overview of the TREC 2023 Deep Learning Track}. In \bibinfo{booktitle}{\emph{Text REtrieval Conference (TREC)}}. NIST, \bibinfo{publisher}{TREC}.
\newblock
\urldef\tempurl%
\url{https://www.microsoft.com/en-us/research/publication/overview-of-the-trec-2023-deep-learning-track/}
\showURL{%
\tempurl}


\bibitem[Faggioli et~al\mbox{.}(2023)]%
        {faggioli2023perspectives}
\bibfield{author}{\bibinfo{person}{Guglielmo Faggioli}, \bibinfo{person}{Laura Dietz}, \bibinfo{person}{Charles~LA Clarke}, \bibinfo{person}{Gianluca Demartini}, \bibinfo{person}{Matthias Hagen}, \bibinfo{person}{Claudia Hauff}, \bibinfo{person}{Noriko Kando}, \bibinfo{person}{Evangelos Kanoulas}, \bibinfo{person}{Martin Potthast}, \bibinfo{person}{Benno Stein}, {et~al\mbox{.}}} \bibinfo{year}{2023}\natexlab{}.
\newblock \showarticletitle{Perspectives on large language models for relevance judgment}. In \bibinfo{booktitle}{\emph{Proceedings of the 2023 ACM SIGIR International Conference on Theory of Information Retrieval}}. \bibinfo{pages}{39--50}.
\newblock


\bibitem[Jayasinghe et~al\mbox{.}(2014)]%
        {jayasinghe2014improving}
\bibfield{author}{\bibinfo{person}{Gaya~K Jayasinghe}, \bibinfo{person}{William Webber}, \bibinfo{person}{Mark Sanderson}, {and} \bibinfo{person}{J~Shane Culpepper}.} \bibinfo{year}{2014}\natexlab{}.
\newblock \showarticletitle{Improving test collection pools with machine learning}. In \bibinfo{booktitle}{\emph{Proceedings of the 19th Australasian Document Computing Symposium}}. \bibinfo{pages}{2--9}.
\newblock


\bibitem[Kaszkiel and Zobel(1997)]%
        {marcin1997passage}
\bibfield{author}{\bibinfo{person}{Marcin Kaszkiel} {and} \bibinfo{person}{Justin Zobel}.} \bibinfo{year}{1997}\natexlab{}.
\newblock \showarticletitle{Passage Retrieval Revisited}. In \bibinfo{booktitle}{\emph{Proceedings of SIGIR}}.
\newblock


\bibitem[Liu et~al\mbox{.}(2023)]%
        {liu2023g}
\bibfield{author}{\bibinfo{person}{Yang Liu}, \bibinfo{person}{Dan Iter}, \bibinfo{person}{Yichong Xu}, \bibinfo{person}{Shuohang Wang}, \bibinfo{person}{Ruochen Xu}, {and} \bibinfo{person}{Chenguang Zhu}.} \bibinfo{year}{2023}\natexlab{}.
\newblock \showarticletitle{G-Eval: NLG Evaluation using Gpt-4 with Better Human Alignment}. In \bibinfo{booktitle}{\emph{The 2023 Conference on Empirical Methods in Natural Language Processing}}.
\newblock


\bibitem[MacAvaney et~al\mbox{.}(2021)]%
        {macavaney2021simplified}
\bibfield{author}{\bibinfo{person}{Sean MacAvaney}, \bibinfo{person}{Andrew Yates}, \bibinfo{person}{Sergey Feldman}, \bibinfo{person}{Doug Downey}, \bibinfo{person}{Arman Cohan}, {and} \bibinfo{person}{Nazli Goharian}.} \bibinfo{year}{2021}\natexlab{}.
\newblock \showarticletitle{Simplified data wrangling with ir\_datasets}. In \bibinfo{booktitle}{\emph{Proceedings of the 44th International ACM SIGIR Conference on Research and Development in Information Retrieval}}. \bibinfo{pages}{2429--2436}.
\newblock


\bibitem[Nguyen et~al\mbox{.}(2016)]%
        {tri2016msmarco}
\bibfield{author}{\bibinfo{person}{Tri Nguyen}, \bibinfo{person}{Mir Rosenberg}, \bibinfo{person}{Xia Song}, \bibinfo{person}{Jianfeng Gao}, \bibinfo{person}{Saurabh Tiwary}, \bibinfo{person}{Rangan Majumder}, {and} \bibinfo{person}{Li Deng}.} \bibinfo{year}{2016}\natexlab{}.
\newblock \showarticletitle{{MS} {MARCO:} {A} Human Generated MAchine Reading COmprehension Dataset}. In \bibinfo{booktitle}{\emph{Proceedings of the Workshop on Cognitive Computation: Integrating neural and symbolic approaches 2016 co-located with the 30th Annual Conference on Neural Information Processing Systems {(NIPS} 2016), Barcelona, Spain, December 9, 2016}}.
\newblock


\bibitem[Rahmani et~al\mbox{.}(2024a)]%
        {rahmani2024synthetic}
\bibfield{author}{\bibinfo{person}{Hossein~A Rahmani}, \bibinfo{person}{Nick Craswell}, \bibinfo{person}{Emine Yilmaz}, \bibinfo{person}{Bhaskar Mitra}, {and} \bibinfo{person}{Daniel Campos}.} \bibinfo{year}{2024}\natexlab{a}.
\newblock \showarticletitle{Synthetic Test Collections for Retrieval Evaluation}.
\newblock \bibinfo{journal}{\emph{arXiv preprint arXiv:2405.07767}} (\bibinfo{year}{2024}).
\newblock


\bibitem[Rahmani et~al\mbox{.}(2024b)]%
        {rahmani2024report}
\bibfield{author}{\bibinfo{person}{Hossein~A Rahmani}, \bibinfo{person}{Clemencia Siro}, \bibinfo{person}{Mohammad Aliannejadi}, \bibinfo{person}{Nick Craswell}, \bibinfo{person}{Charles~LA Clarke}, \bibinfo{person}{Guglielmo Faggioli}, \bibinfo{person}{Bhaskar Mitra}, \bibinfo{person}{Paul Thomas}, {and} \bibinfo{person}{Emine Yilmaz}.} \bibinfo{year}{2024}\natexlab{b}.
\newblock \showarticletitle{Report on the 1st Workshop on Large Language Model for Evaluation in Information Retrieval (LLM4Eval 2024) at SIGIR 2024}.
\newblock \bibinfo{journal}{\emph{arXiv preprint arXiv:2408.05388}} (\bibinfo{year}{2024}).
\newblock


\bibitem[Rahmani et~al\mbox{.}(2024c)]%
        {rahmani2024llm4eval}
\bibfield{author}{\bibinfo{person}{Hossein~A. Rahmani}, \bibinfo{person}{Clemencia Siro}, \bibinfo{person}{Mohammad Aliannejadi}, \bibinfo{person}{Nick Craswell}, \bibinfo{person}{Charles L.~A. Clarke}, \bibinfo{person}{Guglielmo Faggioli}, \bibinfo{person}{Bhaskar Mitra}, \bibinfo{person}{Paul Thomas}, {and} \bibinfo{person}{Emine Yilmaz}.} \bibinfo{year}{2024}\natexlab{c}.
\newblock \showarticletitle{LLM4Eval: Large Language Model for Evaluation in IR}. In \bibinfo{booktitle}{\emph{Proceedings of the 47th International ACM SIGIR Conference on Research and Development in Information Retrieval}} (Washington DC, USA) \emph{(\bibinfo{series}{SIGIR '24})}. \bibinfo{publisher}{Association for Computing Machinery}, \bibinfo{address}{New York, NY, USA}, \bibinfo{pages}{3040–3043}.
\newblock
\showISBNx{9798400704314}
\urldef\tempurl%
\url{https://doi.org/10.1145/3626772.3657992}
\showDOI{\tempurl}


\bibitem[Rahmani et~al\mbox{.}(2024d)]%
        {rahmani2024llmjudge}
\bibfield{author}{\bibinfo{person}{Hossein~A Rahmani}, \bibinfo{person}{Emine Yilmaz}, \bibinfo{person}{Nick Craswell}, \bibinfo{person}{Bhaskar Mitra}, \bibinfo{person}{Paul Thomas}, \bibinfo{person}{Charles~LA Clarke}, \bibinfo{person}{Mohammad Aliannejadi}, \bibinfo{person}{Clemencia Siro}, {and} \bibinfo{person}{Guglielmo Faggioli}.} \bibinfo{year}{2024}\natexlab{d}.
\newblock \showarticletitle{LLMJudge: LLMs for Relevance Judgments}.
\newblock \bibinfo{journal}{\emph{arXiv preprint arXiv:2408.08896}} (\bibinfo{year}{2024}).
\newblock


\bibitem[Sanderson et~al\mbox{.}(2010)]%
        {sanderson2010test}
\bibfield{author}{\bibinfo{person}{Mark Sanderson} {et~al\mbox{.}}} \bibinfo{year}{2010}\natexlab{}.
\newblock \showarticletitle{Test collection based evaluation of information retrieval systems}.
\newblock \bibinfo{journal}{\emph{Foundations and Trends{\textregistered} in Information Retrieval}} \bibinfo{volume}{4}, \bibinfo{number}{4} (\bibinfo{year}{2010}), \bibinfo{pages}{247--375}.
\newblock


\bibitem[Thakur et~al\mbox{.}(2021)]%
        {thakur2021beir}
\bibfield{author}{\bibinfo{person}{Nandan Thakur}, \bibinfo{person}{Nils Reimers}, \bibinfo{person}{Andreas R{\"u}ckl{\'e}}, \bibinfo{person}{Abhishek Srivastava}, {and} \bibinfo{person}{Iryna Gurevych}.} \bibinfo{year}{2021}\natexlab{}.
\newblock \showarticletitle{BEIR: A Heterogeneous Benchmark for Zero-shot Evaluation of Information Retrieval Models}. In \bibinfo{booktitle}{\emph{Thirty-fifth Conference on Neural Information Processing Systems Datasets and Benchmarks Track (Round 2)}}.
\newblock


\bibitem[Thomas et~al\mbox{.}(2023)]%
        {thomas2023large}
\bibfield{author}{\bibinfo{person}{Paul Thomas}, \bibinfo{person}{Seth Spielman}, \bibinfo{person}{Nick Craswell}, {and} \bibinfo{person}{Bhaskar Mitra}.} \bibinfo{year}{2023}\natexlab{}.
\newblock \showarticletitle{Large language models can accurately predict searcher preferences}.
\newblock \bibinfo{journal}{\emph{arXiv preprint arXiv:2309.10621}} (\bibinfo{year}{2023}).
\newblock


\bibitem[Voorhees(2005)]%
        {voorhees2005trec}
\bibfield{author}{\bibinfo{person}{Ellen~M Voorhees}.} \bibinfo{year}{2005}\natexlab{}.
\newblock \showarticletitle{The TREC robust retrieval track}. In \bibinfo{booktitle}{\emph{ACM SIGIR Forum}}, Vol.~\bibinfo{volume}{39}. ACM New York, NY, USA, \bibinfo{pages}{11--20}.
\newblock


\bibitem[Wade and Allan(2005)]%
        {wade2005passage}
\bibfield{author}{\bibinfo{person}{Courtney Wade} {and} \bibinfo{person}{James Allan}.} \bibinfo{year}{2005}\natexlab{}.
\newblock \showarticletitle{Passage retrieval and evaluation}.
\newblock \bibinfo{journal}{\emph{Tech. Reports of DTIC}} (\bibinfo{year}{2005}).
\newblock


\bibitem[Wang et~al\mbox{.}(2023)]%
        {wang2023query2doc}
\bibfield{author}{\bibinfo{person}{Liang Wang}, \bibinfo{person}{Nan Yang}, {and} \bibinfo{person}{Furu Wei}.} \bibinfo{year}{2023}\natexlab{}.
\newblock \showarticletitle{Query2doc: Query Expansion with Large Language Models}. In \bibinfo{booktitle}{\emph{Proceedings of the 2023 Conference on Empirical Methods in Natural Language Processing}}. \bibinfo{pages}{9414--9423}.
\newblock


\bibitem[Zhao et~al\mbox{.}(2024)]%
        {zhao2024dense}
\bibfield{author}{\bibinfo{person}{Wayne~Xin Zhao}, \bibinfo{person}{Jing Liu}, \bibinfo{person}{Ruiyang Ren}, {and} \bibinfo{person}{Ji-Rong Wen}.} \bibinfo{year}{2024}\natexlab{}.
\newblock \showarticletitle{Dense text retrieval based on pretrained language models: A survey}.
\newblock \bibinfo{journal}{\emph{ACM Transactions on Information Systems}} \bibinfo{volume}{42}, \bibinfo{number}{4} (\bibinfo{year}{2024}), \bibinfo{pages}{1--60}.
\newblock


\end{thebibliography}

\newpage

\appendix
\section{Resource Evaluation Results}

\begin{figure*}[!ht]
    \centering
    \subfloat[mAP\label{fig:dl-19-map}]
    {
        {
            \includegraphics[scale=0.31]{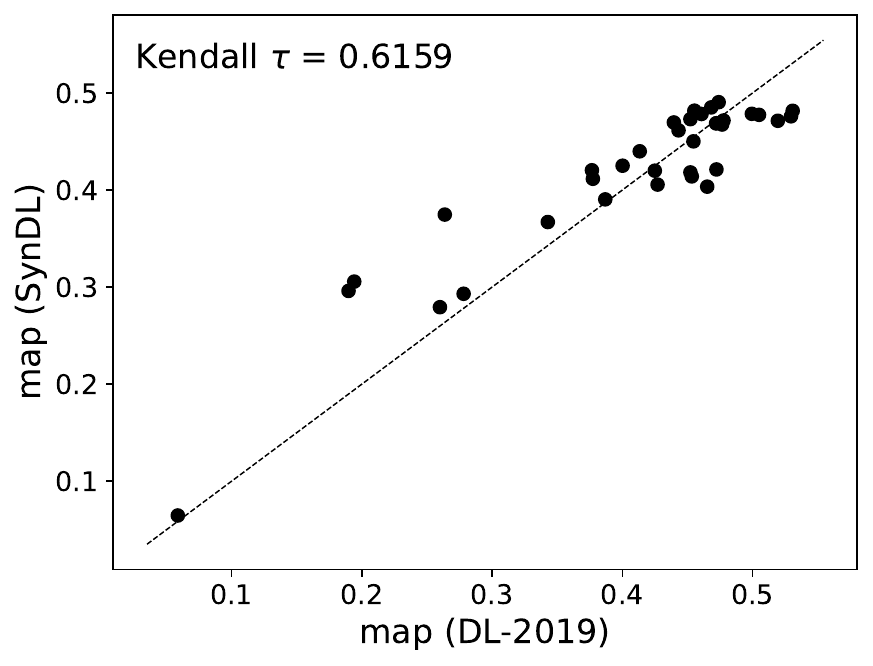}
        }
    }%
    % \quad
    \subfloat[NDCG@5\label{fig:dl-19-ndcg5}]
    {
        {
            \includegraphics[scale=0.31]{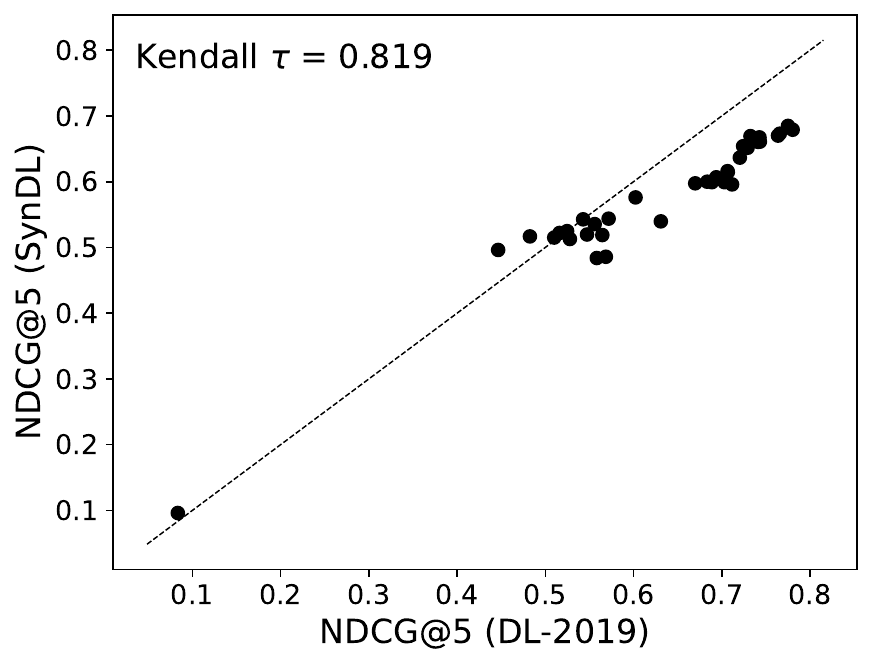}
        }
    }%
    % \quad
    \subfloat[NDCG@10\label{fig:dl-19-ndcg10-app}]
    {
        {
            \includegraphics[scale=0.31]{figs/dl2019/dl2019_NDCG10_black.pdf}
        }
    }%
    % \quad
    \subfloat[NDCG@100\label{fig:dl-19-ndcg100-app}]
    {
        {
            \includegraphics[scale=0.31]{figs/dl2019/dl2019_NDCG100_black.pdf}
        }
    }%
    \caption{System Ranking correlation test between DL-2019 and \dataset.}%
    \label{fig:dl-19-all}%
\end{figure*}

\begin{figure*}
    \centering
    \subfloat[mAP\label{fig:dl-20-map}]
    {
        {
            \includegraphics[scale=0.31]{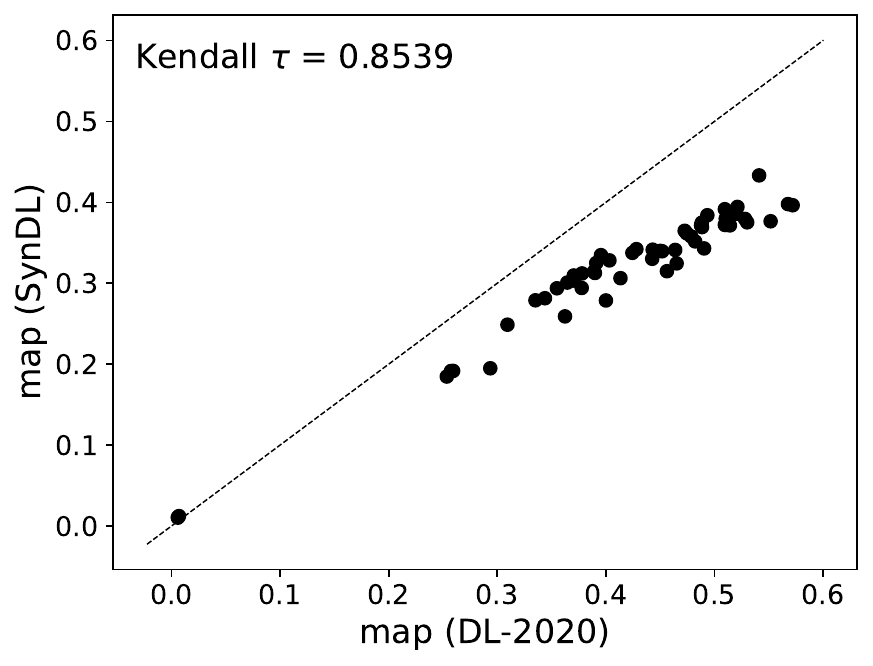}
        }
    }%
    % \quad
    \subfloat[NDCG@5\label{fig:dl-20-ndcg5}]
    {
        {
            \includegraphics[scale=0.31]{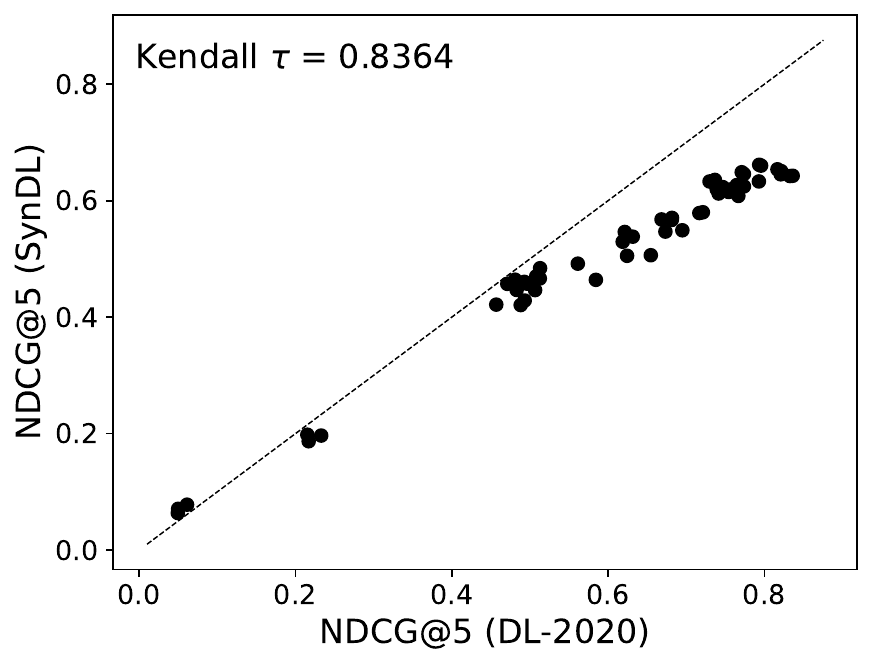}
        }
    }%
    % \quad
    \subfloat[NDCG@10\label{fig:dl-20-ndcg10}]
    {
        {
            \includegraphics[scale=0.31]{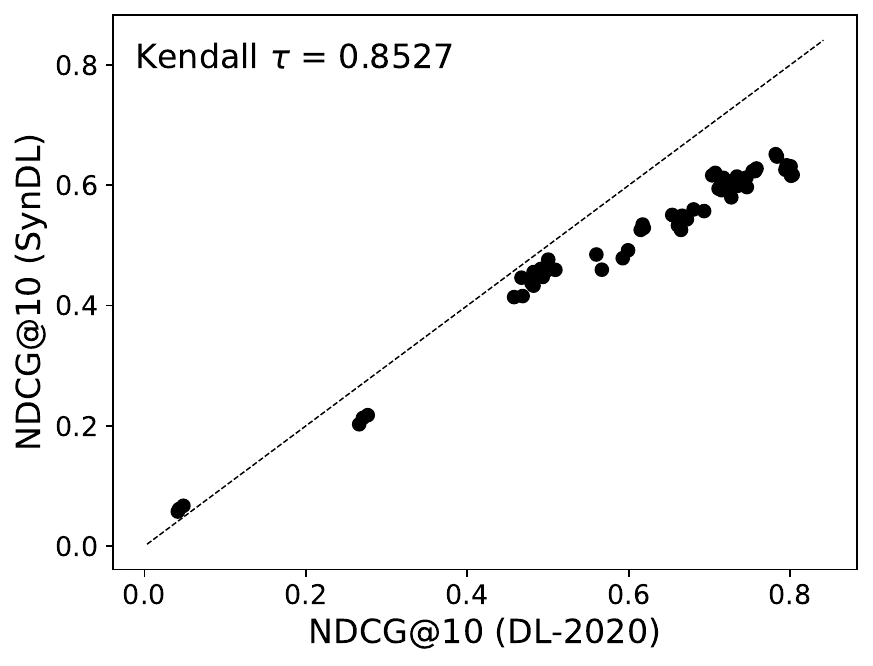}
        }
    }%
    % \quad
    \subfloat[NDCG@100\label{fig:dl-20-ndcg100}]
    {
        {
            \includegraphics[scale=0.31]{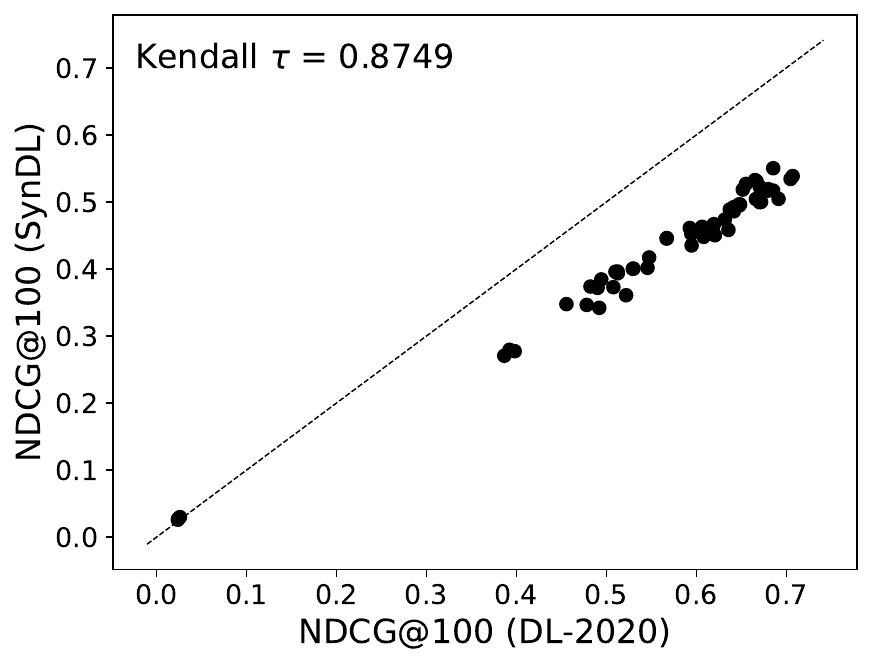}
        }
    }%
    \caption{System Ranking correlation test between DL-2020 and \dataset.}%
    \label{fig:dl-20-all}%
\end{figure*}

\begin{figure*}
    \centering
    \subfloat[mAP\label{fig:dl-21-map}]
    {
        {
            \includegraphics[scale=0.31]{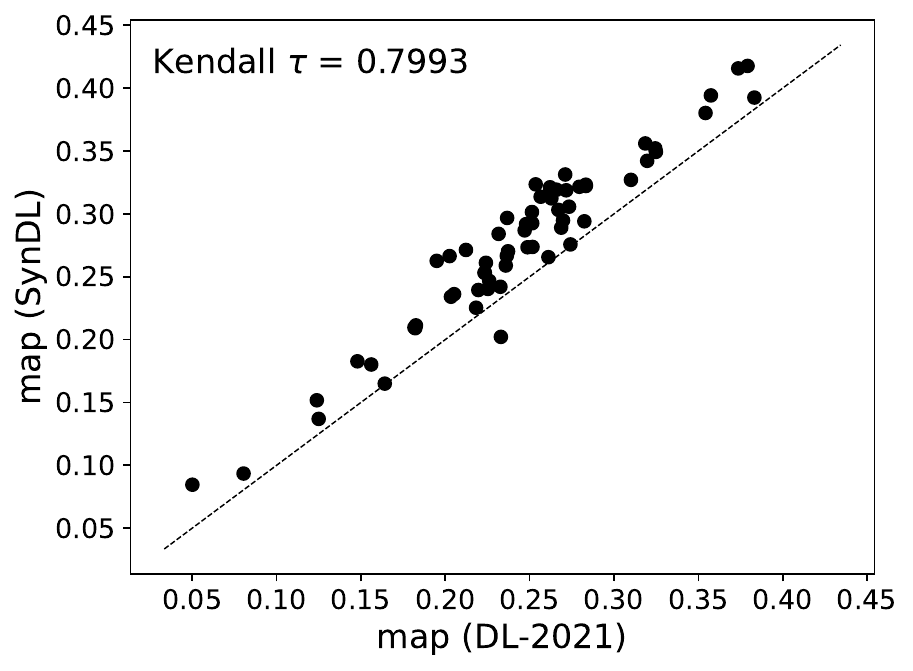}
        }
    }%
    % \quad
    \subfloat[NDCG@5\label{fig:dl-21-ndcg5}]
    {
        {
            \includegraphics[scale=0.31]{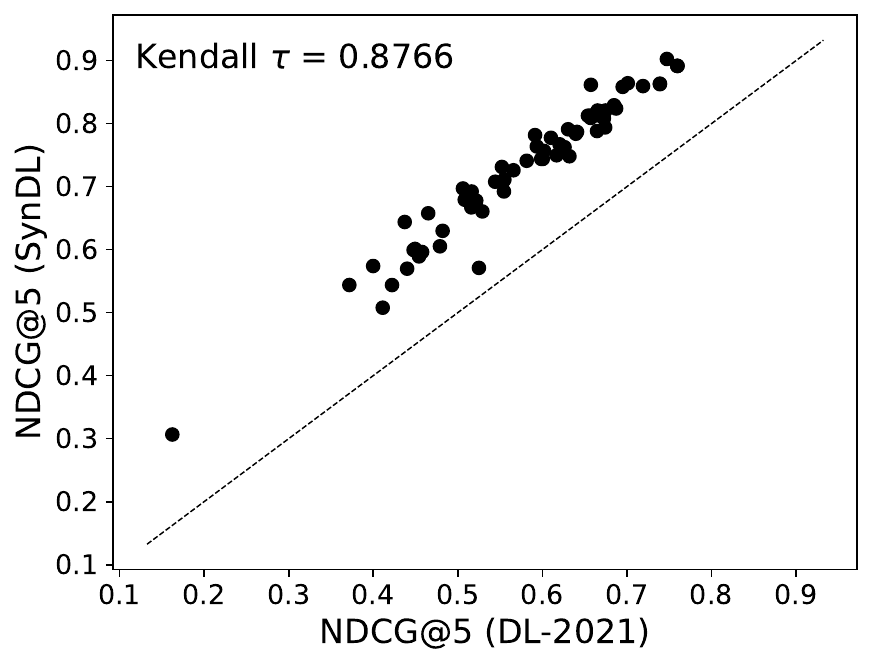}
        }
    }%
    % \quad
    \subfloat[NDCG@10\label{fig:dl-21-ndcg10}]
    {
        {
            \includegraphics[scale=0.31]{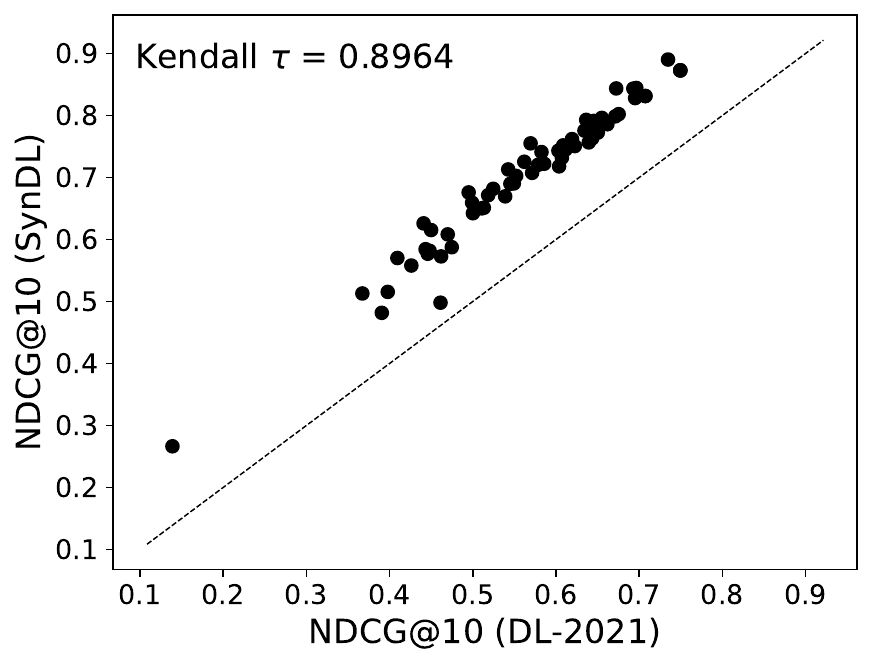}
        }
    }%
    % \quad
    \subfloat[NDCG@100\label{fig:dl-21-ndcg100}]
    {
        {
            \includegraphics[scale=0.31]{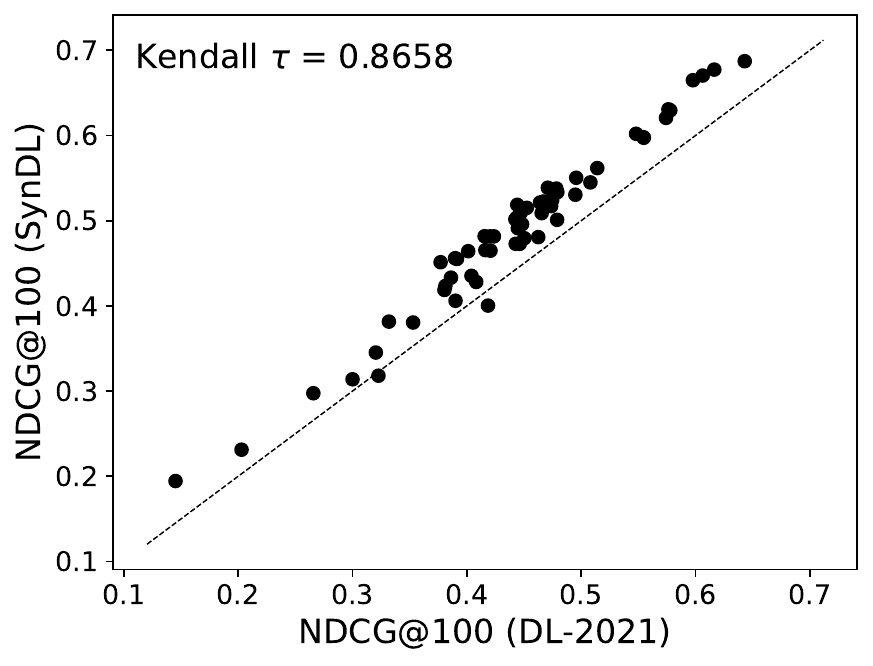}
        }
    }%
    \caption{System Ranking correlation test between DL-2021 and \dataset.}%
    \label{fig:dl-21-all}%
\end{figure*}

\begin{figure*}
    \centering
    \subfloat[mAP\label{fig:dl-22-map}]
    {
        {
            \includegraphics[scale=0.31]{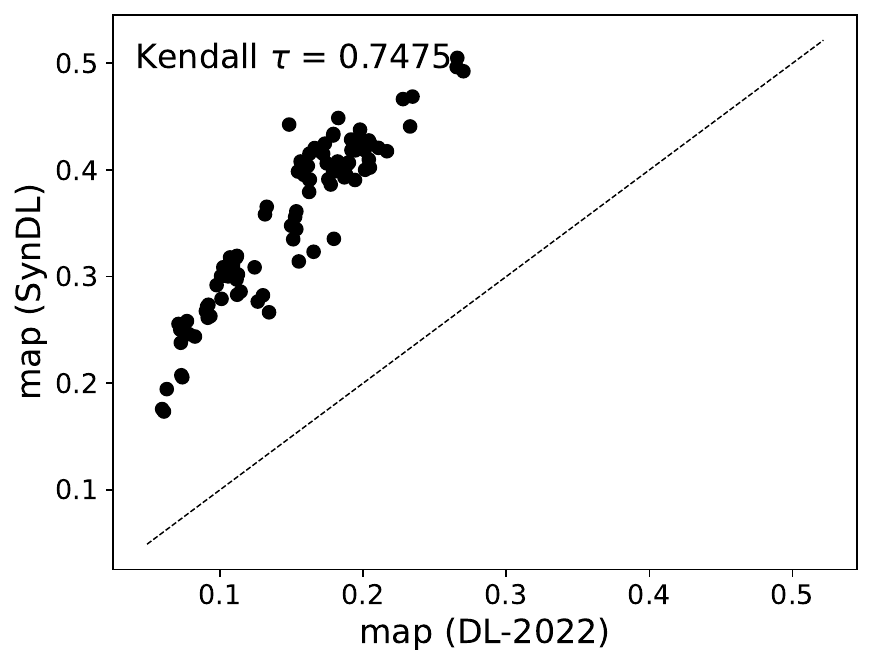}
        }
    }%
    % \quad
    \subfloat[NDCG@5\label{fig:dl-22-ndcg5}]
    {
        {
            \includegraphics[scale=0.31]{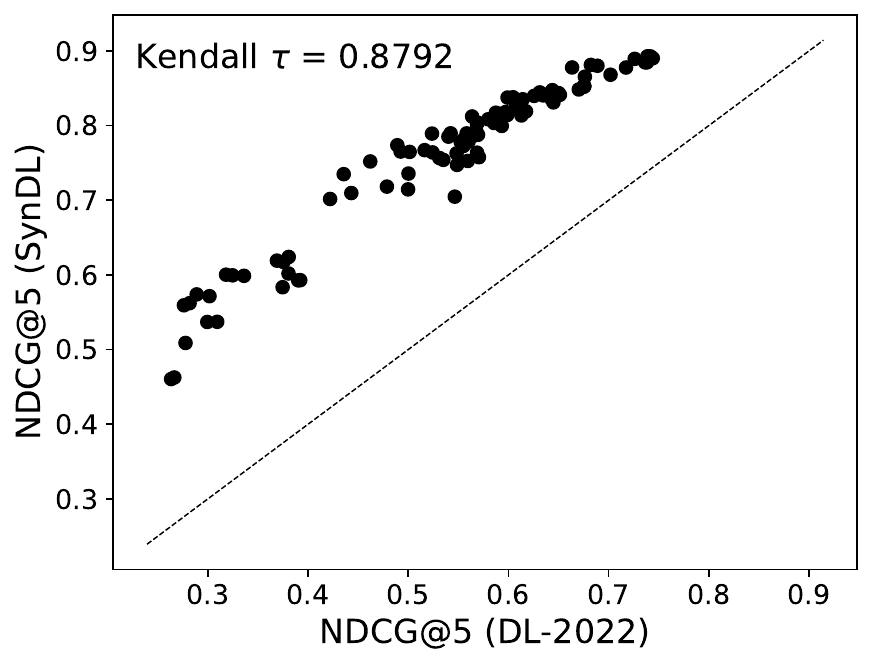}
        }
    }%
    % \quad
    \subfloat[NDCG@10\label{fig:dl-22-ndcg10}]
    {
        {
            \includegraphics[scale=0.31]{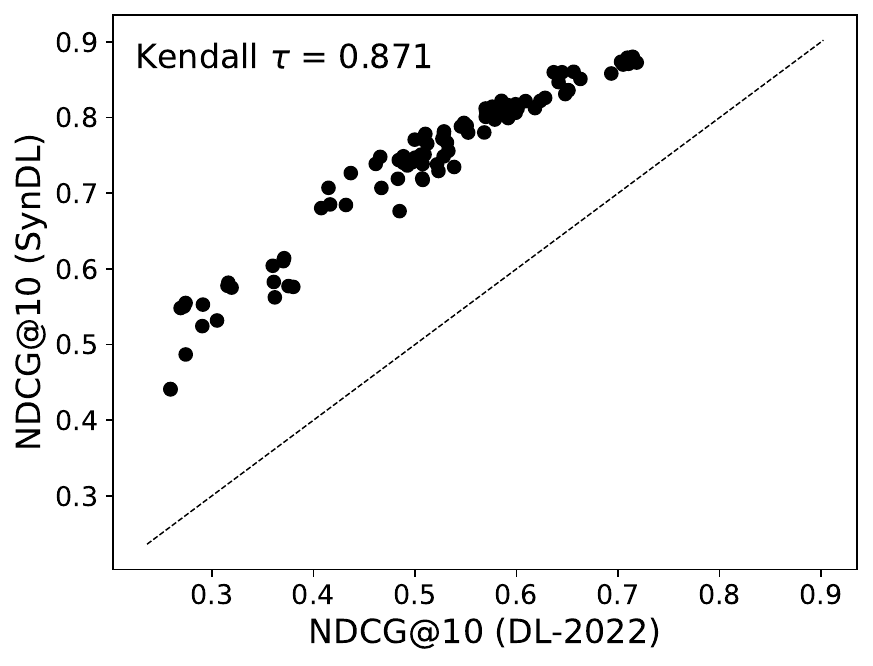}
        }
    }%
    % \quad
    \subfloat[NDCG@100\label{fig:dl-22-ndcg100}]
    {
        {
            \includegraphics[scale=0.31]{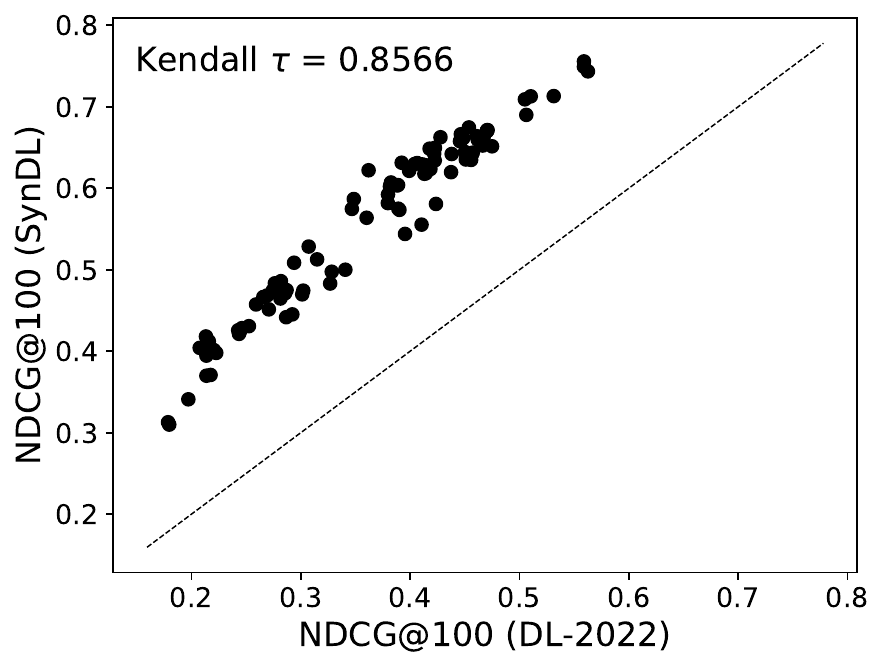}
        }
    }%
    \caption{System Ranking correlation test between DL-2022 and \dataset.}%
    \label{fig:dl-22-all}%
\end{figure*}

\begin{figure*}
    \centering
    \subfloat[mAP\label{fig:dl-23-map}]
    {
        {
            \includegraphics[scale=0.31]{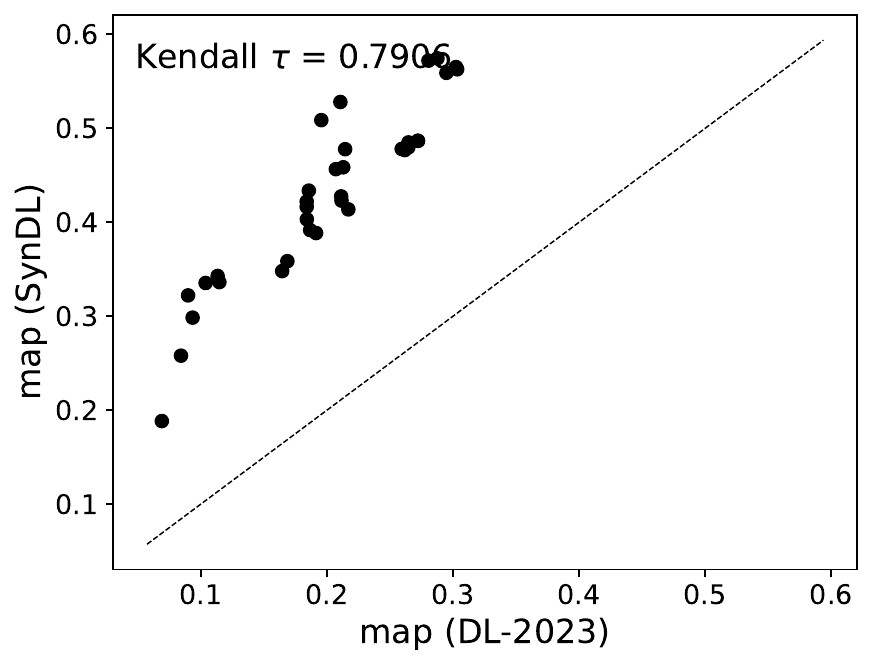}
        }
    }%
    % \quad
    \subfloat[NDCG@5\label{fig:dl-23-ndcg5}]
    {
        {
            \includegraphics[scale=0.31]{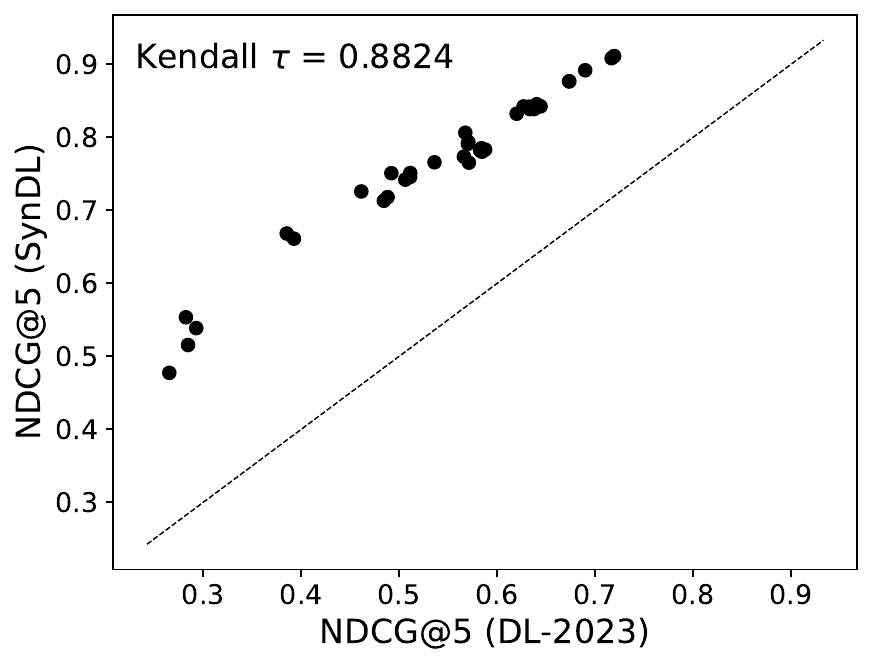}
        }
    }%
    % \quad
    \subfloat[NDCG@10\label{fig:dl-23-ndcg10}]
    {
        {
            \includegraphics[scale=0.31]{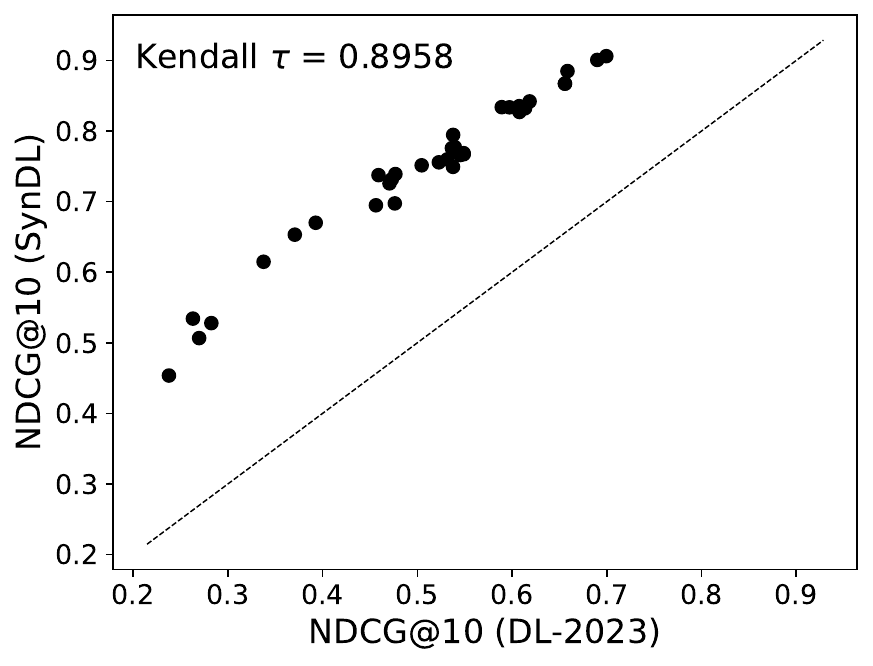}
        }
    }%
    % \quad
    \subfloat[NDCG@100\label{fig:dl-23-ndcg100}]
    {
        {
            \includegraphics[scale=0.31]{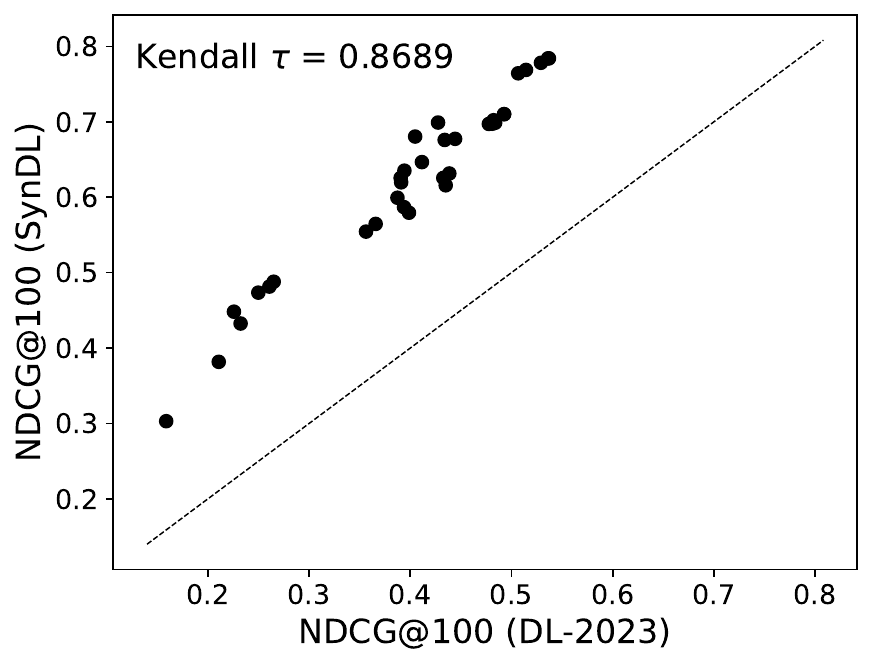}
        }
    }%
    \caption{System Ranking correlation test between DL-2023 and \dataset.}%
    \label{fig:dl-23-all}%
\end{figure*}

\begin{figure*}
    \centering
    \subfloat[mAP\label{fig:dl-23-map-color}]
    {
        {
            \includegraphics[scale=0.31]{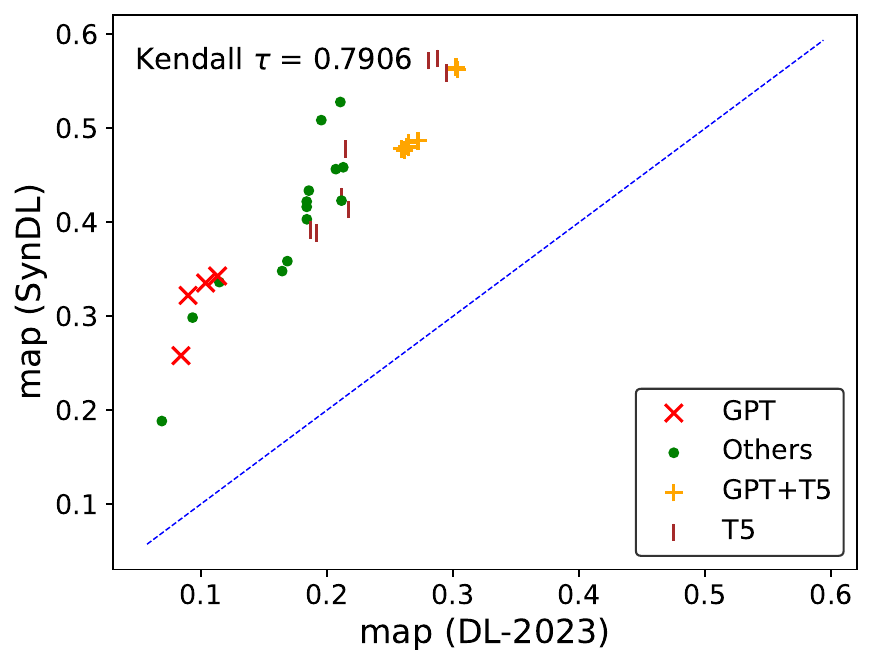}
        }
    }%
    % \quad
    \subfloat[NDCG@5\label{fig:dl-23-ndcg5-color}]
    {
        {
            \includegraphics[scale=0.31]{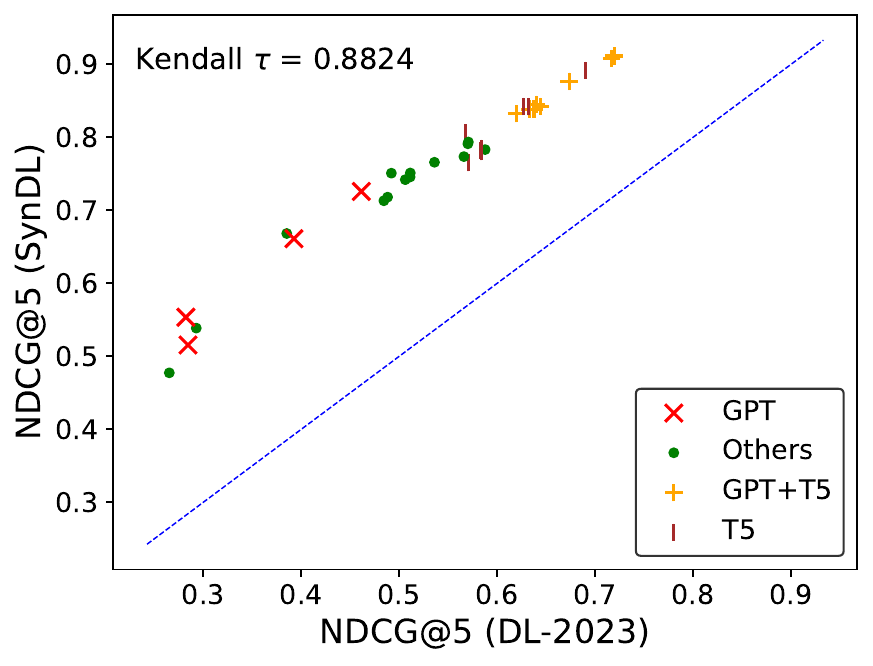}
        }
    }%
    % \quad
    \subfloat[NDCG@10\label{fig:dl-23-ndcg10-color}]
    {
        {
            \includegraphics[scale=0.31]{figs/dl2023/dl2023_NDCG10_colored.pdf}
        }
    }%
    % \quad
    \subfloat[NDCG@100\label{fig:dl-23-ndcg100-color}]
    {
        {
            \includegraphics[scale=0.31]{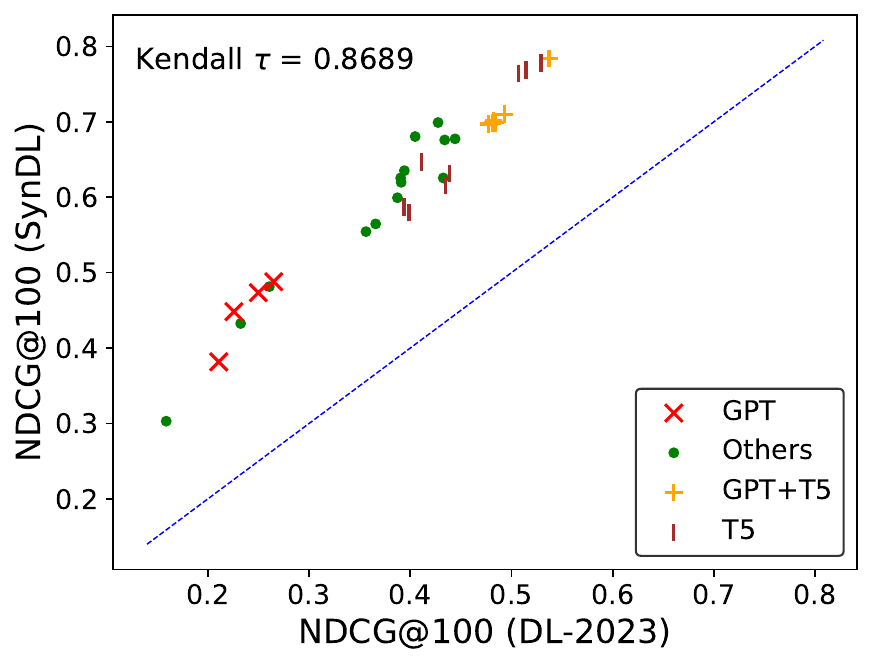}
        }
    }%
    \caption{Scatter plots of the effectiveness of DL-2023 runs based on \dataset synthetic queries vs.~DL-2023 test collection to analyse the bias towards systems using the same language model as the one used in synthetic query construction.}%
    \label{fig:dl-22-colors}%
\end{figure*}

\begin{table*}
    \centering
    \caption{Five top-performing submission runs for each year of DL track and \dataset ($\nearrow$ sorted based on the NDCG@10)}
    \label{tbl:numrels_perquery_dls}
    \begin{tabular}{lcccclcccc}
        \toprule
        \textbf{Run} & \textbf{NDCG@5} & \textbf{NDCG@10$\nearrow$} & \textbf{NDCG@100} & \textbf{mAP} && \textbf{NDCG@5} & \textbf{NDCG@10} & \textbf{NDCG@100} & \textbf{mAP} \\
        \midrule
         & \multicolumn{4}{c}{\textbf{DL-2019}} && \multicolumn{4}{c}{\textbf{\dataset}} \\
        \midrule
        idst\_bert\_p2 & 0.775 & 0.7632 & 0.6828 & 0.5193 && 0.6849 & 0.6771 & 0.6117 & 0.4713 \\
        idst\_bert\_p3 & 0.7803 & 0.7594 & 0.6862 & 0.5307 && 0.679 & 0.6771 & 0.6152 & 0.4816 \\
        p\_exp\_rm3\_bert & 0.7427 & 0.7422 & 0.6745 & 0.5294 && 0.6674 & 0.6611 & 0.6065 & 0.4759 \\
        p\_bert & 0.7334 & 0.738 & 0.6585 & 0.5049 && 0.6637 & 0.656 & 0.6036 & 0.4774 \\
        idst\_bert\_pr2 & 0.7637 & 0.7379 & 0.6357 & 0.4776 && 0.67 & 0.6595 & 0.5916 & 0.4716 \\
        \midrule
         & \multicolumn{4}{c}{\textbf{DL-2020}} && \multicolumn{4}{c}{\textbf{\dataset}} \\
        \midrule
        pash\_r3 & 0.8362 & 0.8031 & 0.6719 & 0.5097 && 0.6428 & 0.6173 & 0.5 & 0.3722 \\
        pash\_r2 & 0.8327 & 0.8011 & 0.6703 & 0.5102 && 0.6424 & 0.6159 & 0.4995 & 0.372 \\
        pash\_f3 & 0.8215 & 0.8005 & 0.6787 & 0.5203 && 0.6508 & 0.6316 & 0.5166 & 0.3859 \\
        pash\_f1 & 0.8168 & 0.7956 & 0.6756 & 0.519 && 0.6539 & 0.6332 & 0.5171 & 0.3857 \\
        pash\_f2 & 0.8209 & 0.7941 & 0.6704 & 0.5105 && 0.6453 & 0.6259 & 0.5092 & 0.3798 \\
        \midrule
         & \multicolumn{4}{c}{\textbf{DL-2021}} && \multicolumn{4}{c}{\textbf{\dataset}} \\
        \midrule
        pash\_f1 & 0.7596 & 0.7494 & 0.5742 & 0.3542 && 0.8915 & 0.8728 & 0.6207 & 0.3802 \\
        pash\_f2 & 0.7596 & 0.7494 & 0.6063 & 0.3736 && 0.8915 & 0.8728 & 0.6704 & 0.4157 \\
        pash\_f3 & 0.7596 & 0.7494 & 0.6164 & 0.3792 && 0.8915 & 0.8728 & 0.6773 & 0.4177 \\
        NLE\_P\_v1 & 0.7472 & 0.7347 & 0.643 & 0.3832 && 0.9025 & 0.8903 & 0.6872 & 0.3925 \\
        pash\_r2 & 0.739 & 0.7076 & 0.4669 & 0.2833 && 0.8625 & 0.8314 & 0.5224 & 0.3232 \\
        \midrule
         & \multicolumn{4}{c}{\textbf{DL-2022}} && \multicolumn{4}{c}{\textbf{\dataset}} \\
        \midrule
        pass3 & 0.7441 & 0.7184 & 0.5313 & 0.2327 && 0.8903 & 0.8727 & 0.7131 & 0.4408 \\
        NLE\_SPLADE\_CBERT\_DT5\_RR & 0.742 & 0.7145 & 0.559 & 0.2657 && 0.8926 & 0.8803 & 0.7557 & 0.5051 \\
        NLE\_SPLADE\_CBERT\_RR & 0.7405 & 0.7141 & 0.559 & 0.2653 && 0.8928 & 0.88 & 0.7491 & 0.4965 \\
        pass2 & 0.7382 & 0.7105 & 0.5103 & 0.2344 && 0.8847 & 0.8711 & 0.7128 & 0.4688 \\
        NLE\_SPLADE\_RR & 0.7387 & 0.7092 & 0.5626 & 0.27 && 0.8928 & 0.8791 & 0.7435 & 0.4926 \\
        \midrule
         & \multicolumn{4}{c}{\textbf{DL-2023}} && \multicolumn{4}{c}{\textbf{\dataset}} \\
        \midrule
        naverloo\/-rgpt4 & 0.7193 & 0.6994 & 0.537 & 0.3032 && 0.9111 & 0.906 & 0.7841 & 0.5628 \\
        naverloo\/-frgpt4 & 0.7166 & 0.6899 & 0.5362 & 0.3023 && 0.9081 & 0.9007 & 0.7841 & 0.5651 \\
        naverloo\_fs\_RR\_duo & 0.6897 & 0.6585 & 0.5291 & 0.2947 && 0.8916 & 0.8849 & 0.7782 & 0.559 \\
        cip\_run\_2 & 0.6734 & 0.6558 & 0.4927 & 0.2721 && 0.8765 & 0.8671 & 0.7101 & 0.4866 \\
        cip\_run\_1 & 0.6733 & 0.6558 & 0.4927 & 0.2721 && 0.8765 & 0.8671 & 0.7101 & 0.4867 \\
        \bottomrule
    \end{tabular}
\end{table*}

\newpage

% Define a custom tcolorbox style for the LLM prompt
\tcbset{
    colback=gray!10,    % Background color of the box
    colframe=gray!80,   % Frame color of the box
    fonttitle=\bfseries,% Title font
    coltitle=black,     % Title color
    boxrule=0.5mm,      % Box border thickness
    arc=4mm,            % Corner rounding
    outer arc=2mm,      % Outer arc rounding
    boxsep=5pt,         % Box separation from text
    left=4mm,           % Left padding
    right=4mm,          % Right padding
    top=2mm,            % Top padding
    bottom=2mm,         % Bottom padding
}

\newtcolorbox{promptbox}[1][]{title=Judgment Prompt, #1}

\section{Judgment Generation Prompt}

\begin{promptbox}
You are a search quality rater evaluating the relevance of passages. Given a query and a web page, you must provide a score on an integer scale of 0 to 3 with the following meanings: \\

3 = Perfectly relevant: The passage is dedicated to the query and contains the exact answer. \\
2 = Highly relevant: The passage has some answer for the query, but the answer may be a bit unclear, or hidden amongst extraneous information. \\
1 = Related: The passage seems related to the query but does not answer it. \\
0 = Irrelevant: The passage has nothing to do with the query. \\

Assume that you are writing an answer to the query. If the passage seems to be related to the query but does not include any answer to the query, mark it 1. If you would use any of the information contained in the passage in such an asnwer, mark it 2. If the passage is primarily about the query, or contains vital information about the topic, mark it 3. Otherwise, mark it 0. \\

A person has typed [\{\textit{\textcolor{gray}{query}}\}] into a search engine. \\
    
\textbf{Result} \\
Consider the following passage. \\
—BEGIN Passage CONTENT— \\
\{\textit{\textcolor{gray}{passage}}\} \\
—END Passage CONTENT— \\

\textbf{Instructions} \\
Consider the underlying intent of the search, and decide on a final score of the relevancy of query to the passage given the context. \\ \\
\textbf{Score:}
\end{promptbox}

\end{document}